\documentclass[
prb,
amsmath,amssymb,
aps,
reprint,
 amsmath,amssymb,
 aps,
]{revtex4-2}

\usepackage{booktabs} 
\usepackage{siunitx}  
\usepackage{makecell} 
\usepackage{multirow} 
\usepackage{xr-hyper}
\usepackage{hyperref}
\usepackage{float}
\usepackage{threeparttable}
\usepackage{siunitx}
\usepackage{graphicx}
\usepackage{dcolumn}
\usepackage{bm}
\usepackage{hyperref}

\begin{document}

\preprint{APS/123-QED}

\title{OpenCSP: A Deep Learning Framework for Crystal Structure Prediction from Ambient to High Pressure}

\author{Yinan Wang}
\affiliation{Artificial Intelligence for Science Institute,
            No.150 Chengfu Road, Haidian District, 
            Beijing,
            100084, 
            China}
\author{Xiaoyang Wang}
\affiliation{National Key Laboratory of Computational Physics,
  Institute of Applied Physics and Computational Mathematics, Fenghao East Road 2, Beijing 100094, P.R.~China}
\author{Zhenyu Wang}
\affiliation{{Key Laboratory of Material Simulation Methods \& Software of Ministry of Education and State Key Laboratory of Superhard Materials, College of Physics, Jilin University, Changchun 130012, P.R.~China}}
\author{Jing Wu}
\affiliation{{Artificial Intelligence for Science Institute,
            No.150 Chengfu Road, Haidian District, 
            Beijing,
            100084, 
            China}}
\author{Jian Lv}
\thanks{Corresponding author: \href{mailto:lvjian@jlu.edu.cn}{lvjian@jlu.edu.cn}}
\affiliation{Key Laboratory of Material Simulation Methods \& Software of Ministry of Education and State Key Laboratory of Superhard Materials, College of Physics, Jilin University, Changchun 130012, P.R.~China}
\author{Han Wang}
\thanks{Corresponding author: \href{mailto:wang_han@iapcm.ac.cn}{wang\_han@iapcm.ac.cn}}
\affiliation{National Key Laboratory of Computational Physics,
  Institute of Applied Physics and Computational Mathematics, Fenghao East Road 2, Beijing 100094, P.R.~China}
\affiliation{HEDPS, CAPT, College of Engineering, Peking University, Beijing 100871, P.R.~China}

\date{\today}

\begin{abstract}
High-pressure crystal structure prediction (CSP) underpins advances in condensed matter physics, planetary science, and materials discovery. Yet, most large atomistic models are trained on near-ambient, equilibrium data, leading to degraded stress accuracy at tens to hundreds of gigapascals and sparse coverage of pressure-stabilized stoichiometries and dense coordination motifs.
Here, we introduce OpenCSP, a machine learning framework for CSP tasks spanning ambient to high-pressure conditions. 
This framework comprises an open-source pressure-resolved dataset alongside a suite of publicly available atomistic models that are jointly optimized for accuracy in energy, force, and stress predictions.
The dataset is constructed via randomized high-pressure sampling and iteratively refined through an uncertainty-guided concurrent learning strategy, which enriches underrepresented compression regimes while suppressing redundant DFT labeling. 
Despite employing a training corpus one to two orders of magnitude smaller than those of leading large models, OpenCSP achieves comparable or superior performance in high-pressure enthalpy ranking and stability prediction.
Across benchmark CSP tasks spanning a wide pressure window, our models match or surpass MACE-MPA-0, MatterSim v1 5M, and GRACE-2L-OAM, with the largest gains observed at elevated pressures. 
These results demonstrate that targeted, pressure-aware data acquisition coupled with scalable architectures enables data-efficient, high-fidelity CSP, paving the way for autonomous materials discovery under ambient and extreme conditions.
\end{abstract}

\maketitle


\par

\section{Introduction}

Pressure is a powerful thermodynamic variable that can dramatically transform materials by shortening interatomic separations, altering electronic structure, and reshaping chemical bonding networks~\cite{zhang2017materials}. 
These effects could drive pressure-induced phase transitions and phenomena absent at ambient conditions, advancing condensed-matter and planetary science while enabling the discovery of superhard, catalytic, optoelectronic, thermoelectric, and superconducting materials~\cite{mao2016recent,xu2022materials,wang2025advances}. 
Two representative cases illustrate this impact: (i) ultrahard diamond and related metastable carbon allotropes, formed only under extreme pressure–temperature conditions yet quenchable to ambient with exceptional mechanical performance due to large kinetic barriers~\cite{huang2014nanotwinned}; 
and (ii) high critical temperature ($T_c \approx 260$ K) superconductivity in the clathrate lanthanum superhydride LaH$_{10}$ near 200~GPa — first predicted and later experimentally confirmed~\cite{liu2017potential,peng2017hydrogen,drozdov2019superconductivity,somayazulu2019evidence} — showing that pressure can stabilize hydrogen-rich frameworks with strongly enhanced electron–phonon coupling. 
These examples underscore the dual role of high-pressure science in enabling technologically relevant metastable phases and revealing quantum states otherwise inaccessible.

Experimentally, high pressures are accessed by (i) static compression in diamond anvil cells (DACs), which now reach multi-megabar (near-TPa) conditions but are limited by microscopic sample volumes, stress gradients, and restricted in situ characterization, and (ii) dynamic (shock, laser, or ramp) compression, which attains similar or higher pressures on nanosecond–microsecond timescales but produces transient, often heated, non-equilibrium states~\cite{jayaraman1983diamond,dubrovinskaia2016terapascal,jeanloz2007achieving}. These practical constraints amplify the need for predictive theoretical guidance, making crystal structure prediction (CSP) an essential complement for high-pressure materials discovery and experimental interpretation~\cite{wang2014perspective,oganov2019structure,wang2022crystal}. CSP aims at finding the ground-state and low-enthalpy metastable structures at specified composition and pressure by globally exploring a high-dimensional enthalpy landscape. This is commonly achieved by coupling density functional theory (DFT) with advanced sampling/optimization strategies such as random structure search (AIRSS)~\cite{pickard2011ab}, particle-swarm optimization (CALYPSO)~\cite{CALYPSO,CALYPSO2}, evolutionary algorithms USPEX~\cite{glass2006uspex} and XtalOpt~\cite{lonie2011xtalopt}, machine-learning accelerated graph-theoretical structure search (MAGUS)~\cite{wang2023magus}, metadynamics~\cite{barducci2011metadynamics} and minima hopping~\cite{goedecker2004minima}. 
These approaches have successfully predicted and guided the realization of notable high-pressure phases.
For example, the semiconducting \textit{oC}40 phase of lithium was first identified computationally~\cite{lv2011predicted} prior to experimental confirmation, and cubic gauche polymeric nitrogen (cg-N), initially proposed from first-principles calculations~\cite{mailhiot1992polymeric}, was later synthesized near 110~GPa and $\sim$2000~K~\cite{2004Single}. Despite these advances, CSP remains computationally expensive. Exhaustive exploration across stoichiometries and cell sizes, along with the need for highly converged enthalpy (energy + $PV$) rankings, necessitates large numbers of iterative DFT relaxations, which sustains a central bottleneck in high-pressure materials discovery.

In recent years, machine learning interatomic potentials (MLPs) have emerged as powerful tools for atomistic simulation, retaining near–DFT accuracy while reducing computational cost by several orders of magnitude. This capability shows great promise for scaling CSP workflows far beyond the limits imposed by expensive electronic-structure calculations~\cite{wang2024concurrent}.
Building upon earlier system- or chemistry-specific potentials, a new generation of broadly trained MLP frameworks, often referred to as universal MLPs, has recently emerged, including DPA~\cite{DPA1,zhang2024dpa,zhang2025graphneuralnetworkera}, M3GNet~\cite{chen2022universal}, CHGNet~\cite{deng2023chgnet}, MACE~\cite{batatia2023foundation}, GNoME-NequIP~\cite{merchant2023scaling}, EquiformerV2-OMat24~\cite{barroso2024open}, Orb v3~\cite{rhodes2025orb}, SevenNet~\cite{kim2024data}, MatterSim~\cite{yang2024mattersim}, and GRACE~\cite{bochkarev2024graph}. These large atomistic models (LAMs) leverage expansive multi-source datasets, equivariant graph neural architectures, and multi-task or multi-fidelity training strategies to improve generalization across diverse chemistries, bonding topologies, coordination environments and distortion regimes.

Leveraging these advances, large-scale discovery frameworks have begun to integrate such models into broader autonomous pipelines. The Graph Networks for Materials Exploration (GNoME) platform combines iterative structure generation, uncertainty-driven active learning, and high-throughput DFT validation to expand the thermodynamic convex hull by millions of candidate crystal structures~\cite{merchant2023scaling}. In parallel, generative approaches such as MatterGen employ diffusion-based sampling guided by stability-aware scoring to propose synthesizable inorganic materials within user-targeted property windows~\cite{zeni2025generative}. Collectively, these developments are establishing an emerging ecosystem in which universal MLPs, active-learning–driven exploration engines, and generative models act synergistically to reduce human effort and accelerate materials discovery and design.

Despite rapid advances brought by LAMs, high‑pressure CSP imposes demands not yet met by existing frameworks. Current training corpora, as exemplified by MPTrj~\cite{deng2023chgnet} and Alexandria~\cite{schmidt2021crystal,schmidt2023machine,wang2023symmetry}, are dominated by near‑equilibrium, ambient‑pressure ($|P| \leq $ a few GPa) structures, leading to degraded stress (virial) fidelity and uncertain transferability when extrapolated to tens or hundreds of gigapascals. 
Accurate enthalpy ranking at high pressure requires simultaneous precision in energy and $PV$ (or full stress tensor) contributions, yet the reported virial errors in many large-scale models remain comparatively high. 
Unconventional stoichiometries, dense coordination motifs, and electronic reorganizations stabilized only under strong compression are systematically underrepresented, impeding discovery of genuinely new high‑pressure phases. 
Although OMat24 broadens configurational diversity through large-scale single point calculations and finite‑temperature (1000–3000~K) molecular dynamic simulations~\cite{barroso2024open}, it lacks explicit high‑pressure sampling; thermal excursions do not substitute for volumetric collapse and coordination changes driven by pressure. 
MatterSim extends nominal coverage to 1000 GPa~\cite{yang2024mattersim}, but at the cost of an enormous dataset ($\approx 1.7 \times 10^7$ configurations) and massive models comprising $4.55 \times 10^7$ parameters. 
Additionally, the training dataset is proprietary, which precludes further data and model development based on this dataset.
A pressure-resolved CALYPSO-derived set spanning 0–300 GPa (657,000 structures)~\cite{luo2024deep} offers explicit pressure labels but inherits a compositional bias toward previously studied systems, leaving broad chemical spaces unsampled. 
Moreover, prevailing active learning strategies rarely aim to reduce uncertainty in pressure or enthalpy.
Several deficiencies - such as limited high-pressure coverage, underrepresentation of pressure-stabilized chemistries, and suboptimal stress accuracy - underscore the need for new specialized datasets that are pressure-aware, uncertainty-guided, and chemically expansive.
Such datasets should be coupled with explicitly optimized data-efficient models to achieve joint precision in predicting energy, force, and stress in the context of high-pressure CSP.

To address these challenges, we introduce OpenCSP, an open-source dataset specifically designed for CSP tasks across ambient to high-pressure conditions, along with a suite of potential energy models trained on this dataset.
Two design choices distinguish OpenCSP from prior efforts in the construction of LAMs. 
First, initial structures are generated via fully randomized CALYPSO sampling without predefined compositional, symmetry, or structural constraints, preserving access to pressure‑stabilized stoichiometries, dense coordinations, and atypical bonding motifs that heuristic filters or ambient‑pressure priors might exclude. Second, a pressure‑aware DP‑GEN concurrent learning strategy~\cite{zhang2020dpgen,wang2024concurrent} adaptively explores both relaxation pathways and converged minima, selectively enriching underrepresented high‑pressure regions of configuration space while suppressing redundant DFT evaluations. This yields a curated corpus of roughly 1.5 million DFT‑labeled configurations — an order of magnitude smaller than several recent universal datasets — yet spanning a broad, nearly continuous stress distribution essential for accurate virial (and thus enthalpy) prediction.

LAMs trained on the OpenCSP dataset achieve precise pressure control during relaxation and deliver high fidelity in energy, force, and virial, enabling reliable enthalpy ranking across diverse compositions at tens to hundreds of gigapascals. In systematic benchmarks, while using markedly less training data, they match or exceed the performance of models trained on substantially larger datasets (e.g., MACE‑MPA‑0, MatterSim v1 5M, GRACE‑2L‑OAM) in high‑pressure phase prediction, particularly in regimes with small pressure‑dependent enthalpy separations. By combining unbiased high‑pressure structure generation and uncertainty‑guided, data‑efficient sampling within a compact yet pressure‑diverse dataset, OpenCSP provides a practical, high‑accuracy foundation for accelerating autonomous high‑pressure CSP pipelines without inflating computational cost or model scale. As the only currently available open-source platform offering both trained models and a full dataset tailored for high-pressure CSP, OpenCSP establishes a transparent and reproducible benchmark for further research under extreme conditions.

\section{OpenCSP Dataset}

\begin{figure*}
    \centering
    \includegraphics[width=1.0\linewidth]{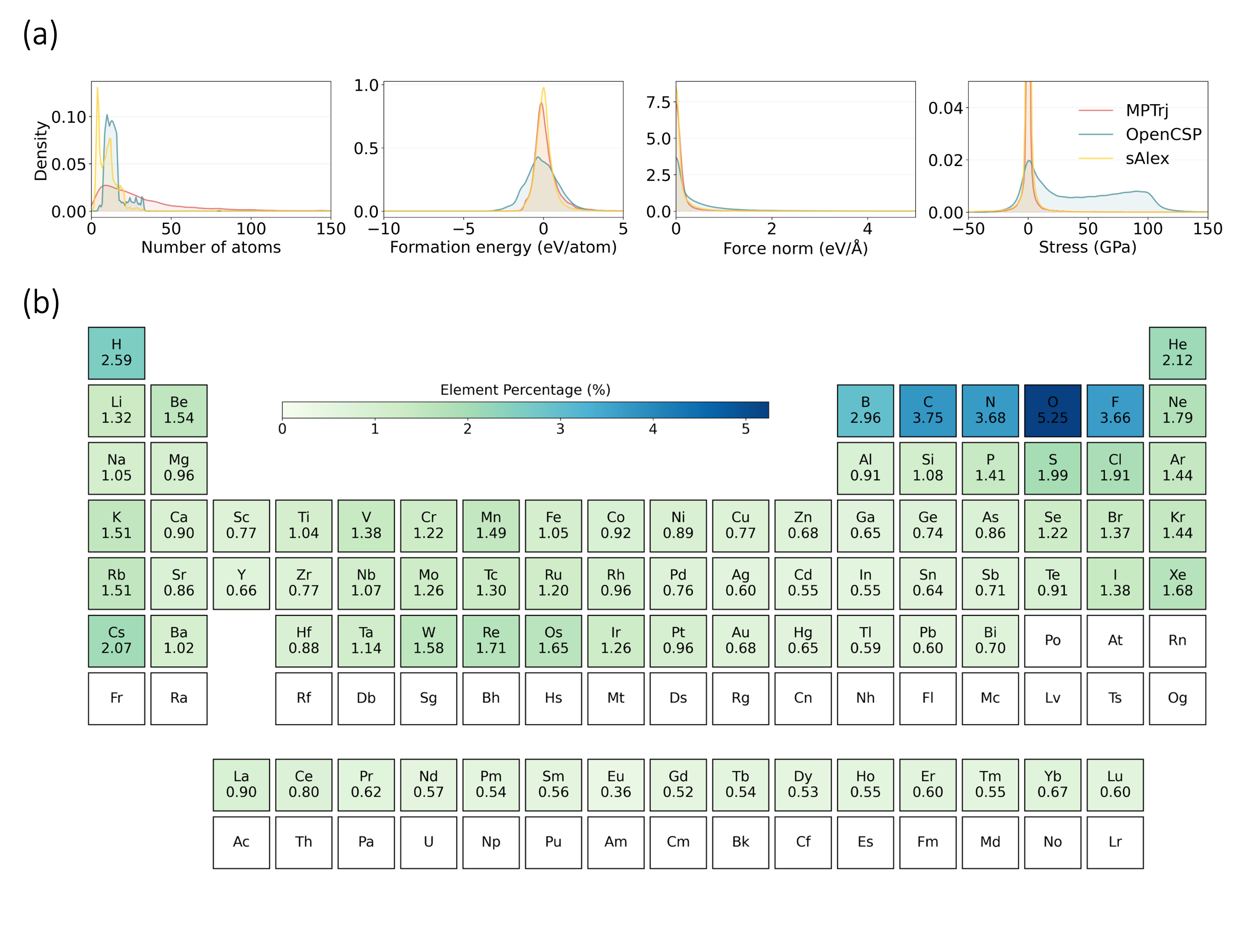}
    \caption{(a) Distributions of number of atoms per structure, formation energy per atom, atomic force norm, and stress (pressure) for the sAlex, MPTrj, and OpenCSP datasets. (b) Elemental occurrence distribution in the OpenCSP dataset.}
    \label{fig:1}
\end{figure*}

The OpenCSP dataset was constructed by relaxing CALYPSO‑proposed structures to pressure‑constrained local minima on the potential energy surface, ensuring direct applicability to CSP tasks. 
All data were obtained through single-point DFT calculations, with structures generated using the DP-GEN concurrent learning framework~\cite{zhang2019active,zhang2020dpgen}, which samples relaxation trajectories under randomly sampled pressure conditions.
For further details, please refer to Section~\ref{sec:dpgen}.
It is important to note that this approach can also incorporate other structure search programs for CSP tasks.

During the course of 113 concurrent learning iterations, a total of 1,522,206 structures were generated, each annotated with DFT-derived energy, force, and virial tensor data. 
The size of the system spans 5 to 32 atoms per cell. 
Although 32 atoms is the upper bound in the training corpus, the resulting models extrapolate reliably to larger cells.
This capability is evidenced by the test results presented in Section~\ref{sec:mptrj}.
During training, the dataset was divided into training, validation, and test sets in an 8:1:1 ratio. 
In particular, the test set contains chemical formulas that are distinct from those present in the training and validation sets.

Figure~\ref{fig:1}(a) presents the distribution of the atom count, formation energy per atom, force norm, and stress (pressure) distribution within the OpenCSP dataset. 
These distributions are compared against those from the MPTrj and sAlex datasets.
In the OpenCSP and sAlex datasets, most configurations contain fewer than 16 atoms. 
In contrast, the MPTrj dataset features a higher proportion of structures with larger atomic counts, although those with fewer than 50 atoms remain predominant.
Across all three datasets the formation energy lies within $-5$ to 5 eV/atom, and most atomic force norms fall below 5 eV/\AA. 
OpenCSP displays a comparatively uniform sampling of stress from about $-10$ to 100 GPa, facilitating model training and evaluation under extreme pressure conditions.
In contrast, the sAlex and MPTrj datasets have stress values concentrated near 0 GPa, indicating a limited presence of high-pressure data.
The OpenCSP dataset encompasses chemical elements from hydrogen (H, atomic number 1) to bismuth (Bi, atomic number 83).
The elemental distribution is depicted in Figure~\ref{fig:1}(b), where a significant overrepresentation of oxygen-containing configurations is evident, a trend also observed in other datasets~\cite{yang2024mattersim,barroso2024open}.

\section{Benchmark of OpenCSP models}

\begin{table*}
\centering
\setlength{\abovecaptionskip}{0.cm}
\setlength{\belowcaptionskip}{0.2cm}
\caption{The information and the size of the training datasets in OpenCSP models and three representative LAMs: MACE‑MPA‑0, GRACE‑2L‑OAM, and MatterSim v1 5M.}
\label{tab:dataset}
\setlength{\tabcolsep}{8pt}{
\begin{threeparttable}
\begin{tabular}{lcc}
\toprule
Models	& Training Dataset &	Data Size \\
\midrule
 OpenCSP-L6 	& OpenCSP 	& 1.5M \\
 OpenCSP-L12 &	 OpenCSP &	 1.5M \\
 OpenCSP-L24 &	 OpenCSP &	 1.5M \\
 MACE-MPA-0 &	 MPTrj+sAlex &	 12M \\
 MatterSim v1 5M &	 MatterSim &	 17M \\
GRACE-2L-OAM &	 OMat24+sAlex+MPTrj 	& 113M \\
\bottomrule
\end{tabular}
\end{threeparttable}}
\end{table*}

Based on the OpenCSP dataset, we trained three models -- OpenCSP‑L6, OpenCSP‑L12, and OpenCSP‑L24 -- implemented with the DPA3 architecture~\cite{zhang2025graphneuralnetworkera}. The suffix denotes the number of stacked interaction (message‑passing) layers (6, 12, and 24, respectively). Training hyperparameters and optimization details are provided in Section~\ref{sec:training}.
To characterize both in‑distribution accuracy and out‑of‑distribution generalization, we conduct five complementary benchmark evaluations:
(i) Accuracy on OpenCSP: energy, force, and virial prediction errors on the training, validation, and held‑out test splits; 
(ii) Cross‑dataset generalization: zero‑shot energy, force, and virial accuracy on the MPTrj dataset; 
(iii) Novel materials formation energy: zero‑shot prediction accuracy for GNoME‑proposed candidate structures;
(iv) Pressure‑controlled structure relaxation: the ability to relax initial structures to specified target pressures (assessment of pressure/volume convergence); and 
(v) CSP under variable pressure: success rates and ranking quality in recovering (or approximating) low‑enthalpy structures across a range of external pressures.
In the first and second case, the virial is defined as the product of pressure and volume of the configuration. 
Virial errors are reported because they offer a direct and unit-consistent measure of the model’s contribution to enthalpy uncertainty.
Consequently, the accuracy of the virial predictions is directly indicative of the accuracy in pressure predictions. 

For comparative evaluation we consider three representative LAMs: MACE‑MPA‑0, GRACE‑2L‑OAM, and MatterSim v1 5M, with details of their respective training datasets provided in Table~\ref{tab:dataset}.
Notably, the OpenCSP models utilize a substantially smaller dataset, approximately an order of magnitude fewer configurations than MACE-MPA-0 and MatterSim v1 5M, and nearly two orders of magnitude fewer than GRACE-2L-OAM.
MatterSim v1 5M is uniquely trained on a dataset comprising 17 million entries produced through molecular dynamics simulations under pressure conditions ranging from 0 to 1000 GPa.
In contrast, MACE-MPA-0 and GRACE-2L-OAM do not explicitly account for pressure effects in the generation of their training data.
Other LAMs, including Orb v3 and SevenNet-MF-ompa~\cite{park2024scalable}, are not examined in our study primarily due to the considerable computational resources required to assess all test cases. 
Additionally, MACE-MPA-0 and GRACE-2L-OAM are considered representative, as they achieve rankings similar to those of Orb v3 and SevenNet-MF-ompa on the Matbench Discovery leaderboard~\cite{riebesell2024matbenchdiscoveryframework}.
The recently released Universal Models for Atoms (UMA)~\cite{wood2025umafamilyuniversalmodels} were not available for evaluation in this study due to licensing constraints.

\subsection{In-distribution accuracy on OpenCSP}\label{sec:dptest}

\begin{table*}
\centering
\setlength{\abovecaptionskip}{0.cm}
\setlength{\belowcaptionskip}{0.2cm}
\caption{Mean absolute error (MAE) and root mean square error (RMSE) for energy (E), force (F), and virial (V) on the OpenCSP training, validation, and test splits. Errors for each property are denoted as $\mathrm{X_{MAE}}$ and $\mathrm{X_{RMSE}}$ (X = E, F, V). All error metrics are reported in their respective units: meV/atom for energy (E), meV/Å~for force (F), and meV/atom for virial (V).}
\label{tab:dptest}
\setlength{\tabcolsep}{1.5mm}{
\begin{threeparttable}
\begin{tabular}{lrrrrrrrrr}
\toprule
& \multicolumn{3}{c}{OpenCSP-L6} & \multicolumn{3}{c}{OpenCSP-L12} & \multicolumn{3}{c}{OpenCSP-L24} \\
\cmidrule(lr){2-4} \cmidrule(lr){5-7} \cmidrule(lr){8-10}
 & {Train} & {Valid} & {Test} & {Train} & {Valid} & {Test} & {Train} & {Valid} & {Test} \\
\midrule
${\mathrm{E_{MAE}}}$ & 21.0 & 25.3 & 26.8 & 16.8 & 22.4 & 23.2 & 13.3 & 19.7 & 20.2 \\
${\mathrm{E_{RMSE}}}$  & 34.9 & 35.2 & 36.9 & 29.9 & 30.9 & 31.8 & 25.9 & 27.3 & 27.9 \\
${\mathrm{F_{MAE}}}$  & 108.0 & 124.0 & 129.0 & 92.9 & 111.0 & 116.7 & 80.1 & 102.0 & 108.0 \\
${\mathrm{F_{RMSE}}}$  & 193.0 & 216.0 & 229.0 & 168.0 & 195.0 & 203.9 & 150.0 & 180.0 & 190.2 \\
${\mathrm{V_{MAE}}}$  & 43.5 & 47.7 & 49.3 & 38.5 & 43.2 & 43.2 & 44.5 & 39.9 & 41.2 \\
${\mathrm{V_{RMSE}}}$ & 87.9 & 88.9 & 105.0 & 79.4 & 81.6 & 81.6 & 82.8 & 75.5 & 78.0 \\
\bottomrule
\end{tabular}
\end{threeparttable}}
\end{table*}

Table~\ref{tab:dptest} summarizes the mean absolute error (MAE) and root mean square error (RMSE) for energy, force, and virial on the OpenCSP training, validation, and test splits; the abbreviations ${\mathrm{E_{MAE}}}$, ${\mathrm{E_{RMSE}}}$, ${\mathrm{F_{MAE}}}$, ${\mathrm{F_{RMSE}}}$, ${\mathrm{V_{MAE}}}$, and ${\mathrm{V_{RMSE}}}$ denote the respective errors for each property.
All three models display strong generalization, as evidenced by the small train-test gaps ($<$ 15\%) and minimal validation-test deviations ($<$ 2.5\%), demonstrating robust predictive capability even for unseen chemical systems.
Notably, increasing network depth consistently improves accuracy, with the 24-layer model achieving the lowest test errors (${\mathrm{E_{MAE}}}$: 20.2 meV/atom; ${\mathrm{F_{MAE}}}$: 108.0 meV/\AA; ${\mathrm{V_{MAE}}}$: 41.2 meV/atom). 
However, the benefits of increasing model depth are highly property-specific.
Energy predictions exhibit sustained improvements, with a 13.4\% reduction in error from 6 to 12 layers and a further 12.9\% reduction from 12 to 24 layers.
In contrast, force and virial predictions show clear diminishing returns beyond 12 layers: force errors decrease by only 7.46\% from 12 to 24 layers (compared to 9.53\% from 6 to 12), while virial improvements drop sharply to 4.63\% (from 12.37\% over the same initial depth increase).
Given these marginal gains relative to the substantial increase in computational cost, deeper architectures were not pursued in this study.

\subsection{Cross-dataset generalization}\label{sec:mptrj}

We assess out‑of‑distribution transfer using the MPTrj dataset, which is a large, trajectory‑level corpus widely used for constructing LAMs. From the MPTrj dataset, we randomly sampled 500 structures containing fewer than 32 atoms and 500 structures containing more than 32 atoms to create the ``MPTrj-S'' and ``MPTrj-L'' test cases, respectively.
Importantly, we deliberately excluded any structures that share the same chemical composition with those in the OpenCSP initial dataset.
MPTrj‑L explicitly probes extrapolation ability to larger systems absent during training.
For the OpenCSP models, the reference energy, force, and virial in the test cases were relabeled using the ABACUS code~\cite{abacus1,abacus2} to ensure alignment with the DFT settings of the training dataset. 
For MACE‑MPA‑0, GRACE‑2L‑OAM, and MatterSim v1 5M we retain the original MPTrj labels, as the labeling method used for their training datasets is consistent with that of the MPTrj dataset.

\begin{table*}
\centering
\setlength{\abovecaptionskip}{0.cm}
\setlength{\belowcaptionskip}{0.2cm}
\caption{Performance comparison on MPTrj test sets partitioned by structure size: MPTrj-S ($<$32 atoms, N=500) and MPTrj-L ($>$32 atoms, N=500). Error metrics in respective units: energy (meV/atom), force (meV/Å), virial (meV/atom).}
\label{tab:mptrj}
\setlength{\tabcolsep}{1.5mm}{
\begin{threeparttable}
\begin{tabular}{lcccccc}
\toprule
\multicolumn{7}{c}{MPTrj-S (structures with $<$32 atoms)} \\
\midrule
Models & OpenCSP-L6 & OpenCSP-L12 & OpenCSP-L24 & MACE-MPA-0 & MatterSim v1 5M & GRACE-2L-OAM \\
\midrule
${\mathrm{E_{MAE}}}$ & 19.1 & 14.8 & 12.7 & 12.0 & 25.1 & 8.6 \\
${\mathrm{E_{RMSE}}}$ & 30.4 & 21.8 & 18.3 & 42.7 & 60.4 & 31.2 \\
${\mathrm{F_{MAE}}}$ & 72.5 & 61.3 & 56.3 & 33.0 & 48.4 & 23.8 \\
${\mathrm{F_{RMSE}}}$ & 133.5 & 123.0 & 110.6 & 64.1 & 104.9 & 55.8 \\
${\mathrm{V_{MAE}}}$ & 29.4 & 24.8 & 22.8 & 167.6 & 158.6 & 160.1 \\
${\mathrm{V_{RMSE}}}$ & 67.9 & 58.6 & 54.5 & 672.7 & 653.6 & 663.3 \\
\midrule
\multicolumn{7}{c}{MPTrj-L (structures with $>$32 atoms) } \\
\midrule
${\mathrm{E_{MAE}}}$ & 15.8 & 13.5 & 10.1 & 6.7 & 13.8 & 6.2 \\
${\mathrm{E_{RMSE}}}$ & 24.4 & 21.3 & 16.0 & 24.6 & 31.9 & 24.6 \\
${\mathrm{F_{MAE}}}$ & 113.3 & 102.0 & 83.3 & 39.9 & 59.7 & 38.6 \\
${\mathrm{F_{RMSE}}}$ & 207.9 & 253.3 & 150.1 & 72.6 & 119.5 & 86.1 \\
${\mathrm{V_{MAE}}}$ & 20.4 & 17.6 & 15.5 & 115.1 & 111.1 & 110.5 \\
${\mathrm{V_{RMSE}}}$ & 40.4 & 37.4 & 33.7 & 533.5 & 527.5 & 533.3 \\
\bottomrule
\end{tabular}
\end{threeparttable}}
\end{table*}

Table~\ref{tab:mptrj} reports MAE and RMSE for energy, force, and virial across all models on MPTrj‑S and MPTrj‑L.  
Within the OpenCSP family, increased depth yields systematically improved accuracy. 
In the MPTrj-S case, OpenCSP-L24 reduces the MAE in energy by 33.5\% compared to OpenCSP-L6, decreasing from 19.1 to 12.7 meV/atom. 
Similarly, force predictions show a 22.3\% reduction in force MAE, and virial predictions see a 22.4\% reduction in virial MAE. 
This trend of improvement is also observed in the MPTrj-L case (36.1\%, 26.5\%, and 24.0\% reduction in energy, force, and virial MAE, respectively), confirming the effectiveness of deeper model.
Particularly, the OpenCSP models demonstrate impressive generalizability to the MPTrj-L case, as the disparity in prediction accuracy between MPTrj-S and MPTrj-L is marginal.
Indeed, the OpenCSP models achieve slightly higher accuracy in energy and virial predictions, despite a significant decrease in force accuracy in the MPTrj-L case, highlighting their robustness across different system sizes.

It is important to note that MPTrj configurations are included in the training datasets of MACE‑MPA‑0 and GRACE‑2L‑OAM, so their reported metrics correspond to (partial) in‑distribution training performance rather than zero‑shot transfer. In contrast, the OpenCSP and MatterSim v1 5M results represent true test (zero‑shot) accuracy. As expected, MACE‑MPA‑0 and GRACE‑2L‑OAM achieve lower energy and force errors than the zero‑shot OpenCSP models. MatterSim v1 5M attains an energy MAE comparable to OpenCSP‑L6 on MPTrj‑S and to OpenCSP‑L12 on MPTrj‑L, while yielding somewhat lower force MAEs. A notable divergence appears in virial prediction: all OpenCSP models produce substantially smaller virial errors than MACE‑MPA‑0, GRACE‑2L‑OAM, and MatterSim v1 5M. Even the shallowest OpenCSP‑L6 model attains a virial MAE only about 18\% of that of the three baselines. Moreover, the virial RMSE/MAE ratio for OpenCSP ($\approx$~2) is markedly lower than the 4–5 range observed for the baseline models, indicating a reduced tail of large virial (PV term) outliers. This suppression of heavy‑tail error is consequential for downstream enthalpy‑sensitive tasks, where occasional large PV deviations can dominate uncertainty.

\subsection{Novel materials formation energy}

The GNoME study reported approximately $3.8\times 10^{5}$ previously unreported crystal structures predicted to lie on the 0 K convex hull of formation energy.
Because these “on‑hull” candidates are absent from all current model training corpora, they constitute a stringent out‑of‑distribution benchmark for assessing a model’s ability to predict the stability of genuinely novel, potentially synthesizable materials. 
To facilitate this evaluation, we randomly subsampled 190 structures from the GNoME reported novel structures and examined the models' precision in predicting their formation energy.
For the OpenCSP models, the structures were relaxed and labeled using ABACUS to ensure compatibility with the DFT settings employed for the training set.
In contrast, the originally reported energy was used for the MACE-MPA-0, MatterSim v1 5M, and GRACE-2L-OAM models, as the DFT settings in GNoME align with the training DFT settings of these models.

\begin{figure}
    \centering
    \includegraphics[width=1\linewidth]{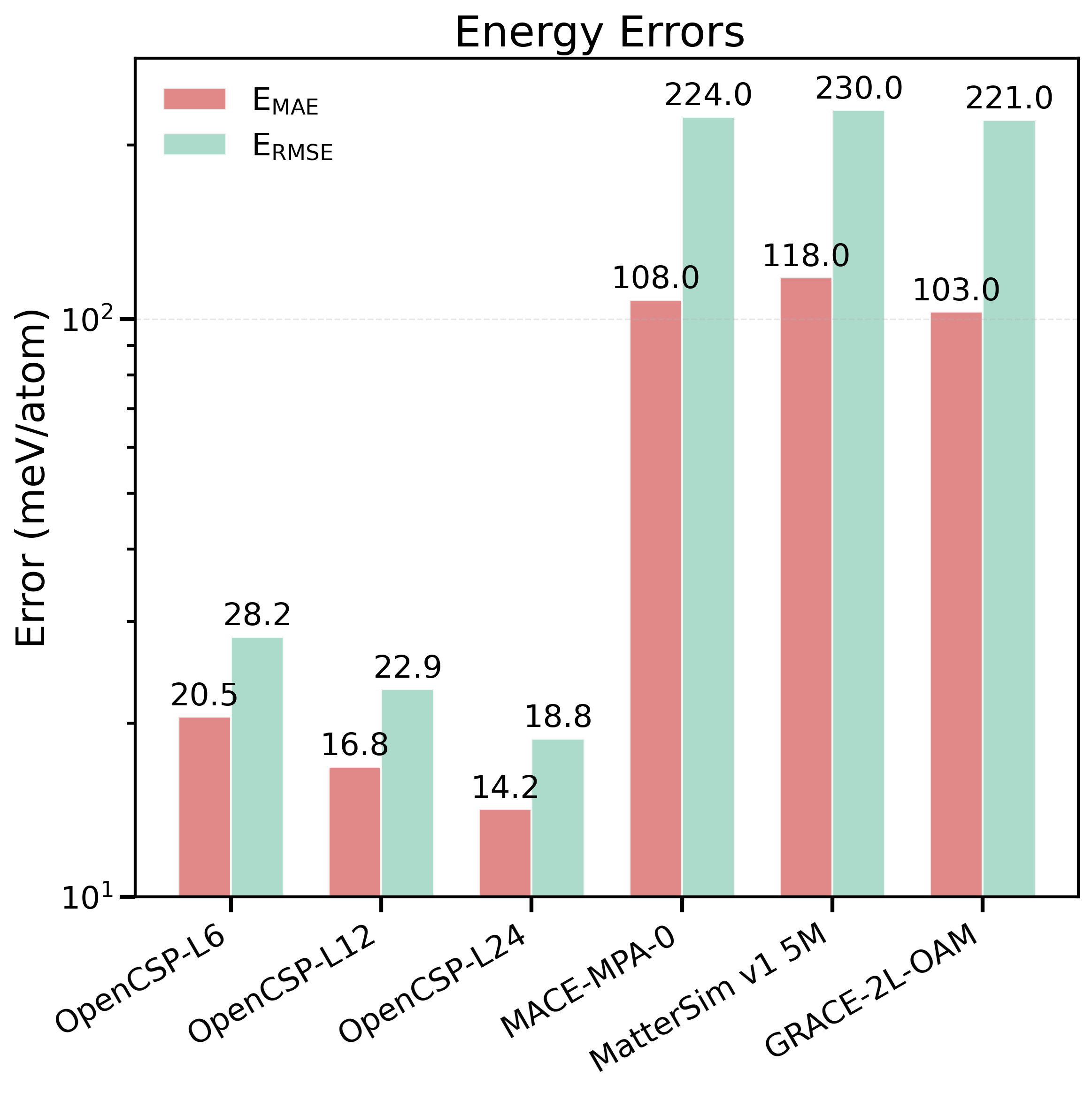}
    \caption{Comparison of formation energy prediction errors (MAE and RMSE) for various models on 190 genuinely novel and stable materials from the GNoME database. }
    \label{fig:GNoME}
\end{figure}

Figure~\ref{fig:GNoME} presents the MAE and RMSEs for the energy predictions of various models. 
The MAEs for the MACE-MPA-0, MatterSim v1 5M, and GRACE-2L-OAM models are approximately 100 meV/atom. 
In contrast, the OpenCSP models exhibit superior accuracy, with MAEs ranging from 20.5 meV/atom for OpenCSP-L6 to 14.2 meV/atom for OpenCSP-L24. 
This result illustrates a clear trend of enhanced generalizability with deeper models, particularly highlighting OpenCSP-L24, which achieves a 30.7\% improvement in precision over OpenCSP-L6.
The evaluation of the formation energy for GNoME on-hull structures underscores the state-of-the-art predictive capability of the OpenCSP models on novel structures compared to existing methods, indicating its significant potential for advancing material discovery.

\subsection{Pressure-controlled structure relaxation}\label{sec:2}

\begin{table*}
\centering
\setlength{\abovecaptionskip}{0.cm}
\setlength{\belowcaptionskip}{0.2cm}
\caption{Reproduction of target external pressures (0, 50, 100, 150, 200 GPa) after model-driven relaxations. For each pressure, 200 randomly generated ternary structures (total N = 1000) were relaxed with each model; the final hydrostatic pressures were then obtained from single-point DFT calculations (ABACUS for OpenCSP relaxations; VASP for MACE-MPA-0, MatterSim v1 5M, and GRACE-2L-OAM). Entries report mean ± standard deviation (GPa).
}
\label{tab:5}
\setlength{\tabcolsep}{3mm}{
\begin{threeparttable}
\begin{tabular}{l *{5}{S[table-format=3.0(4)]}}
\toprule
Model & \multicolumn{1}{c}{0 GPa} & \multicolumn{1}{c}{50 GPa} & \multicolumn{1}{c}{100 GPa} & \multicolumn{1}{c}{150 GPa} & \multicolumn{1}{c}{200 GPa} \\
\midrule
MACE-MPA-0        & 13 \pm 131  & 108 \pm 357 & 208 \pm 397 & 346 \pm 958  & 582 \pm 1824 \\
MatterSim v1 5M   & -1 \pm 24   & 50 \pm 8    & 123 \pm 205 & 146 \pm 13   & 195 \pm 19   \\
GRACE-2L-OAM      & 3 \pm 36    & 90 \pm 287  & 218 \pm 556 & 478 \pm 1200 & 910 \pm 1972 \\
OpenCSP-L6        & 0 \pm 1     & 50 \pm 1    & 100 \pm 2   & 150 \pm 2    & 201 \pm 3    \\
OpenCSP-L12       & 0 \pm 1     & 50 \pm 1    & 100 \pm 1   & 150 \pm 1    & 200 \pm 3    \\
OpenCSP-L24       & 0 \pm 2     & 50 \pm 1    & 100 \pm 1   & 150 \pm 1    & 200 \pm 3    \\
\bottomrule
\end{tabular}
\end{threeparttable}}
\end{table*}

To benchmark the capability of the models in pressure-constrained structural optimization—a core operation in CSP tasks—we generated 1,000 random ternary initial structures (3–30 atoms per cell) using CALYPSO. These were partitioned evenly into five groups and relaxed at external pressures of 0, 50, 100, 150, and 200~GPa, respectively. All relaxations were carried out with the Atomic Simulation Environment (ASE)~\cite{larsen2017atomic}, using identical convergence criteria: a maximum residual atomic force $f_{\max} \le 5.0$~meV/\AA\ and an upper limit of 1,000 optimization steps.

\begin{figure}
    \centering
    \includegraphics[width=1.05\linewidth]{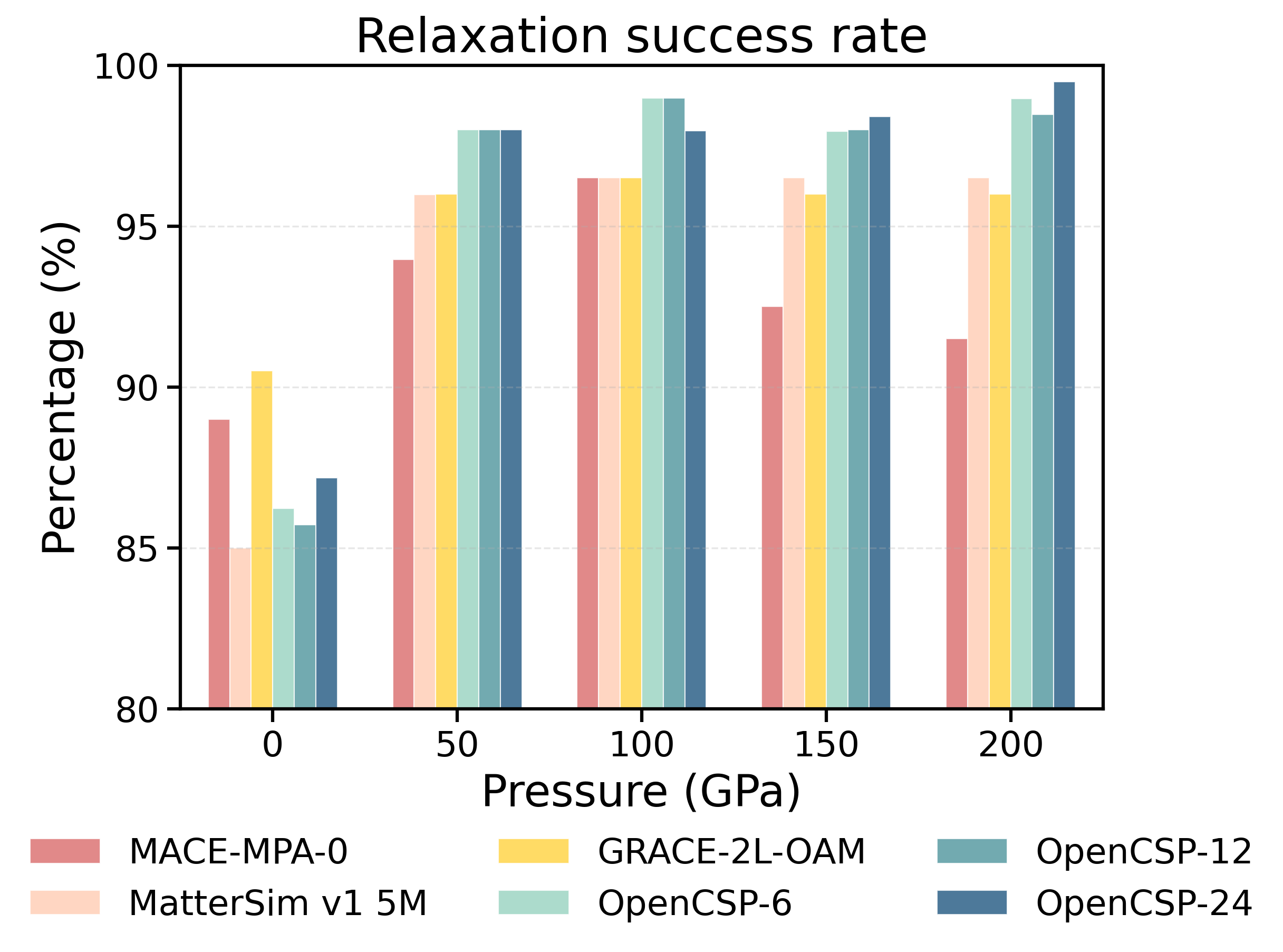}
    \caption{Relaxation success rate for each model at target pressures 0, 50, 100, 150, and 200 GPa. For each pressure, 200 randomly generated ternary structures were relaxed (total N = 1000 per model). A run is counted as successful if the maximum residual force satisfies $f_{\max} \le 5.0$~meV/\AA\ within 1,000 steps.}
    \label{fig:3}
\end{figure}

After relaxation, every final structure was re-evaluated by a single-point DFT   calculation to obtain a consistent, model-independent assessment. Structures relaxed with the OpenCSP models were evaluated using ABACUS, while those relaxed with MACE-MPA-0, MatterSim v1 5M, and GRACE-2L-OAM were evaluated using VASP~\cite{kresse1996efficient,kresse1996efficiency}. The resulting stress tensors were used to quantify agreement with the target pressure. For each model and pressure condition, the mean and standard deviation of the DFT-evaluated pressures were reported in Table~\ref{tab:5}. We define the relaxation success rate as the fraction of cases in which $f_{\max} \le 5.0$ meV/\AA\ was achieved within 1,000 steps (Fig.~\ref{fig:3}).

As summarized in Table~\ref{tab:5}, the OpenCSP models reproduce the imposed external pressures with high fidelity across the full range even at 150 and 200~GPa, conditions beyond their training domain. 
The DFT-evaluated pressure dispersion grows only mildly with load and is still limited to $\approx$3~GPa at 200~GPa ($\approx$1.5 \% relative deviation), indicating stable extrapolation and effective virial control. 
MatterSim v1 5M attains reasonable convergence in most cases, though at a target pressure 100~GPa, the mean pressure shifts to 123~GPa and the larger standard deviation points to reduced pressure consistency relative to OpenCSP. 
MACE-MPA-0 and GRACE-2L-OAM exhibit even more pronounced overpressurization at non-ambient conditions: for a 50~GPa target their mean relaxed pressures reach 108 and 90~GPa, respectively (errors $>$ 80\%). This systematic drift is consistent with the limited coverage of explicit high-pressure sampling and stress-focused training in their source datasets. Even at 0~GPa, their residual pressure spreads (standard deviations of 131 and 36~GPa, respectively) greatly exceed those of the OpenCSP models, underscoring inferior control of the stress tensor. Overall, these comparisons highlight the importance of incorporating pressure-diverse data and virial-aware objectives when targeting reliable enthalpy ranking under compression.

Figure~\ref{fig:3} presents the relaxation success rate. 
At ambient pressure, the GRACE-2L-OAM and MACE-MPA-0 models achieve the highest performance, with success rates of 90.5\% and 89.0\%, respectively. 
These are followed by the OpenCSP and MatterSim v1 5M models, which have success rates ranging from 85\% to 87\%.
As the pressure exceeds 50 GPa, the OpenCSP models demonstrate superior performance, achieving a success rate of over 98\%. 
The success rate of the MACE-MPA-0 model initially increases to 96.5\% at 100 GPa, but then decreases to 91.5\% at 200 GPa. 
The MatterSim v1 5M and GRACE-2L-OAM models exhibit similar behavior, maintaining a success rate of 96.0\% without significant variation at pressures $\geq$50 GPa. 
Nevertheless, for GRACE-2L-OAM and MACE-MPA-0, the high success rates do not coincide with accurate pressure reproduction, as structures that satisfy the convergence criterion can still display substantial residual deviations from the target external pressure. 

\subsection{CSP tasks at different pressure conditions}

\begin{table*}
\centering
\caption{Sources of known structures across pressure ranges used for the high-pressure CSP benchmark.}
\label{tab:pressure_tasks}
\begin{tabular}{c c p{10cm}}
\toprule
\multicolumn{1}{c}{Pressure (GPa)} & \multicolumn{1}{c}{Number of tasks} & \multicolumn{1}{c}{Source} \\
\midrule
0 & 66 & Materials Project on hull structures \\
50 & 8 & ${\mathrm{Li_{12}Au_{4}}}$, ${\mathrm{Li_{1}Au_{1}}}$, ${\mathrm{Li_{20}Au_{4}}}$, ${\mathrm{Li_{2}Au_{1}}}$, ${\mathrm{Li_{8}Au_{2}}}$~\cite{yang2016gold}, ${\mathrm{Mg_{2}Si_{4}O_{16}H_{8}}}$~\cite{shao2024new}, ${\mathrm{Mg_{4}Xe_{2}}}$, ${\mathrm{Mg_{6}Xe_{4}}}$~\cite{miao2015anionic} \\
100 & 12 & ${\mathrm{Cs_{1}F_{3}}}$, ${\mathrm{Cs_{2}F_{6}}}$, ${\mathrm{Cs_{2}F_{8}}}$~\cite{miao2013caesium}, ${\mathrm{Cs_{4}Li_{12}}}$~\cite{botana2014pressure}, ${\mathrm{Al_{12}S_{16}}}$, ${\mathrm{Al_{4}S_{2}}}$, ${\mathrm{Al_{6}S_{6}}}$, ${\mathrm{Al_{8}S_{16}}}$~\cite{shao2020exotically}, ${\mathrm{Mg_{2}Xe_{2}}}$, ${\mathrm{Mg_{2}Xe_{4}}}$, ${\mathrm{Mg_{6}Xe_{2}}}$, ${\mathrm{Mg_{6}Xe_{4}}}$~\cite{miao2015anionic} \\
150 & 11 & ${\mathrm{Cs_{8}F_{40}}}$~\cite{miao2013caesium}, ${\mathrm{Cs_{1}Li_{1}}}$, ${\mathrm{Cs_{2}Li_{10}}}$, ${\mathrm{Cs_{2}Li_{4}}}$, ${\mathrm{Cs_{4}Li_{12}}}$, ${\mathrm{Cs_{4}Li_{16}}}$~\cite{botana2014pressure}, ${\mathrm{Al_{4}S_{6}}}$, ${\mathrm{Al_{4}S_{8}}}$~\cite{shao2020exotically}, ${\mathrm{Xe_{12}N_{4}}}$, ${\mathrm{Xe_{2}N_{10}}}$, ${\mathrm{Xe_{3}N_{18}}}$~\cite{peng2015stable}\\
200 & 23 & ${\mathrm{Al_{18}S_{24}}}$, ${\mathrm{Al_{1}S_{1}}}$, ${\mathrm{Al_{3}S_{6}}}$, ${\mathrm{Al_{8}S_{12}}}$~\cite{shao2020exotically}, ${\mathrm{Cs_{4}Li_{20}}}$~\cite{botana2014pressure}, ${\mathrm{Mg_{10}Xe_{4}}}$, ${\mathrm{Mg_{1}Xe_{1}}}$, ${\mathrm{Mg_{2}Xe_{4}}}$, ${\mathrm{Mg_{4}Xe_{2}}}$, ${\mathrm{Mg_{6}Xe_{2}}}$, ${\mathrm{Mg_{6}Xe_{4}}}$~\cite{miao2015anionic}, ${\mathrm{Xe_{12}N_{16}}}$, ${\mathrm{Xe_{16}N_{12}}}$, ${\mathrm{Xe_{2}N_{8}}}$, ${\mathrm{Xe_{3}N_{1}}}$, ${\mathrm{Xe_{3}N_{2}}}$, ${\mathrm{Xe_{4}N_{28}}}$, ${\mathrm{Xe_{4}N_{32}}}$, ${\mathrm{Xe_{4}N_{4}}}$~\cite{peng2015stable} \\
\bottomrule
\end{tabular}
\end{table*}

\begin{figure}
    \centering    \includegraphics[width=1.05\linewidth]{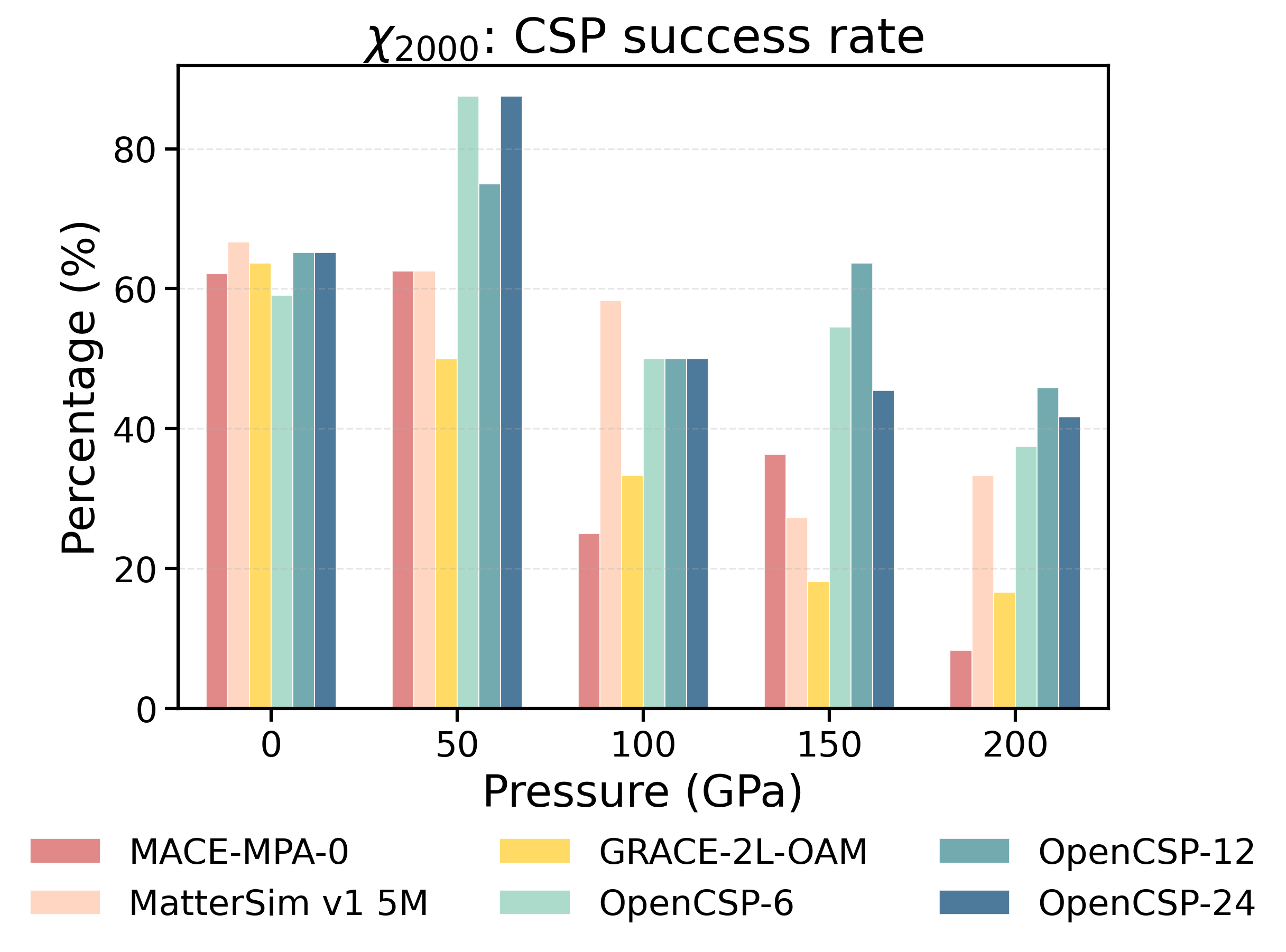}
    \caption{CSP success rate $\chi_{2000}$. For each external pressure, $\chi_{2000}$ is the fraction of benchmark compositions whose reference (on-hull) structure was recovered within 2,000 CALYPSO-generated starting structures.
}
    \label{fig:4}
\end{figure}

To assess each model’s ability to recover thermodynamically stable (on-hull) crystal structures, we evaluated performance in CSP tasks under various external pressures. 
At 0 GPa, we randomly selected 66 on-hull structures from the Materials Project whose chemical compositions are absent from the OpenCSP training dataset.  
For each composition, 2,000 random initial configurations were generated using CALYPSO (fixed stoichiometry, no symmetry constraints) and subsequently relaxed with each model. The relaxed candidates were compared to the corresponding reference structure using pymatgen's \verb|StructureMatcher|~\cite{ong2013python}, with matching tolerances set to \verb|ltol|=0.1 (10\% deviation in lattice parameters), \verb|stol|=0.2 (0.2 Å atomic position tolerance), and \verb|angle_tol|=5 (5-degree angular variation).
We define the CSP success rate, $\chi_{2000}$, as the fraction of compositions for which at least one of the 2,000 relaxed structures matches the reference (66 total compositions at ambient pressure). 
A higher value of $\chi_{2000}$ indicates an increased likelihood of discovering thermodynamically stable structures using the model-driven CSP algorithm within the fixed proposal budget.

Under high-pressure conditions (50, 100, 150, and 200 GPa), reference on-hull structures were collected from the literature , including Li-Au~\cite{yang2016gold}, Cs-F~\cite{miao2013caesium}, Cs-Li~\cite{botana2014pressure}, Al-S~\cite{shao2020exotically}, Mg-Si-O-H~\cite{shao2024new}, Mg-Xe~\cite{miao2015anionic}, and Xe-N~\cite{peng2015stable}. 
For each composition–pressure pair we applied the same protocol as in the ambient-pressure benchmark: generating 2,000 random initial structures for each chemical composition using CALYPSO and relaxing them under target pressures using the models. 
The final relaxed structures were then compared to the literature on-hull configurations, applying an identical structural matching criterion.
The number of benchmark tasks at each pressure, together with source references, is summarized in Table~\ref{tab:pressure_tasks}.

Figure~\ref{fig:4} reports the CSP success rate, $\chi_{2000}$. 
At 0~GPa, all models achieved comparable performances of approximately 60\%, indicating similar baseline capability in recovering previously unseen on-hull structures under ambient conditions.
At high pressures (50-150 GPa), OpenCSP models show substantial improvement, outperforming baseline models by $\approx$ 20-30 percentage points.
Specifically, at 50~GPa, the OpenCSP models exhibit a pronounced improvement, reaching success rates of roughly 80 \%, while the baseline models (MACE-MPA-0, GRACE-2L-OAM, MatterSim v1 5M) do not show a corresponding gain. 
Across 100–150~GPa the OpenCSP success rate stabilizes near 50\%, before declining modestly to about 40\% at 200~GPa - a pressure regime that substantially exceeds the dominant range of explicit sampling in the OpenCSP training data. 
In contrast, MACE-MPA-0, GRACE-2L-OAM, and MatterSim v1 5M display a generally monotonic degradation with increasing pressure. 
The single notable exception is MatterSim v1 5M at 100~GPa (58.3\%), where it surpasses the OpenCSP models, consistent with its inclusion of extensive high-pressure molecular dynamics configurations during training. 
Overall, the pressure-resolved CSP benchmarks highlight the enhanced robustness of the OpenCSP models at elevated pressures (particularly $\geq$50~GPa), underscoring their suitability for accelerating the discovery of novel high-pressure materials.

\section{Conclusion}

We introduced OpenCSP, an open-source high-pressure atomistic dataset containing 1.5 million DFT-labeled configurations. 
These configurations were generated through unbiased CALYPSO sampling integrated with a pressure-aware DP-GEN concurrent learning workflow, covering a pressure range of 0–100 GPa.
We are publicly releasing a suite of three deep learning potential energy models—OpenCSP-L6, OpenCSP-L12, and OpenCSP-L24—that are jointly optimized for predicting energy, force, and virial using the OpenCSP dataset.
Across a broad evaluation suite—including predictions on MPTrj trajectories, generalization to newly reported on-hull structures from GNoME, pressure-constrained structural relaxations, and CSP under varying external pressures—the OpenCSP models deliver consistently strong performance. 
Relative to established baselines (MACE-MPA-0, MatterSim v1 5M, GRACE-2L-OAM), the models—especially the deeper variants—exhibit higher accuracy, improved robustness at elevated pressure, stable convergence in relaxation workflows, and superior CSP success rates.
These results establish OpenCSP as an effective foundation for universal atomistic simulation under variable pressure conditions and a practical platform for accelerated discovery of materials stabilized by compression.
Its unique open availability fills a critical gap in the community, offering researchers a reproducible and transparent benchmark for high-pressure materials modeling.

\section{Method}
\subsection{Data generation: DP-GEN schemes}\label{sec:dpgen}

\begin{figure*}
    \centering
    \includegraphics[width=1.0\linewidth]{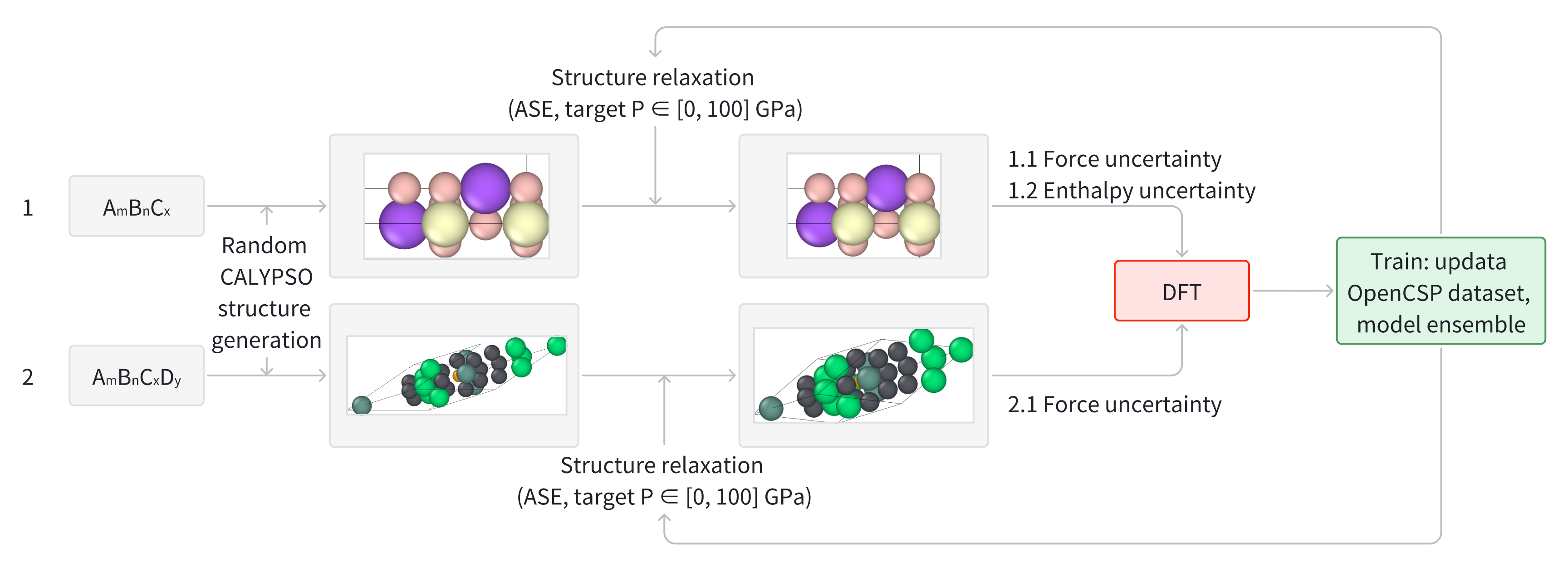}
    \caption{Schematic illustration of the OpenCSP dataset construction workflow based on the concurrent learning scheme.}
    \label{fig:2}
\end{figure*}

The OpenCSP dataset was constructed using an iterative concurrent learning approach~\cite{wang2024concurrent} implemented within the DP-GEN framework. 
Prior to launching the DP-GEN workflow (Figure \ref{fig:2}), an initial dataset was prepared, consisting of four parts:
(1) 610 elemental structures downloaded from the Materials Project~\cite{jain2013commentary}, 
(2) 19,357 bulk trajectory snapshots from the MPTrj dataset, 
(3) 17,468 randomly generated structures from CALYPSO, and 
(4) 326,339 structures from MD simulations on 20,000 selected MPTrj structures.
These structures were labled by DFT calculated energy, atomic force, and virial tensors.

The DP-GEN framework operates as an iterative workflow as depicted in Figure \ref{fig:2}. 
During each iteration, models are trained using the current training dataset (the initial dataset for iteration 1) and then applied in CALYPSO structure prediction tasks to explore the configurational space.
Candidate structures for DFT calculations are selected based on the models' uncertainty estimators.
The labeled structures are then incorporated into the training dataset for the subsequent iteration.
Consequently, each DP-GEN iteration comprises three distinct stages:

(I) Training: 
An ensemble of four models employing the DPA3 architecture~\cite{zhang2025graphneuralnetworkera} was trained using the current dataset.
The training parameters were consistent with those of OpenCSP-L6, as detailed in Sec.~\ref{sec:training}, except that each iteration comprised 100,000 training steps.

\begin{table*}
\centering
\caption{Minimum interatomic distance parameters (\(r_s\)) employed in calculating the safe interatomic distances.}
\label{tab:safeR}
\begin{tabular}{llll}
\toprule
Element & Radius (\AA) & Element & Radius (\AA) \\
\hline
H   & 0.439 & He  & 0.323 \\
Li  & 0.776 & Be  & 0.673 \\
B   & 0.569 & C   & 0.569 \\
N   & 0.569 & O   & 0.569 \\
F   & 0.543 & Ne  & 0.828 \\
Na  & 0.750 & Mg  & 0.879 \\
Al  & 0.879 & Si  & 0.828 \\
P   & 0.776 & S   & 0.776 \\
Cl  & 0.776 & Ar  & 0.984 \\
K   & 1.189 & Ca  & 1.189 \\
Sc  & 1.034 & Ti  & 1.034 \\
V   & 1.034 & Cr  & 0.984 \\
Mn  & 1.009 & Fe  & 0.984 \\
Co  & 0.984 & Ni  & 0.984 \\
Cu  & 0.984 & Zn  & 0.984 \\
Ga  & 1.034 & Ge  & 1.034 \\
As  & 1.034 & Se  & 1.086 \\
Br  & 1.086 & Kr  & 1.189 \\
Rb  & 1.294 & Sr  & 1.294 \\
Y   & 1.086 & Zr  & 1.086 \\
Nb  & 1.086 & Mo  & 1.086 \\
Tc  & 1.086 & Ru  & 1.086 \\
Rh  & 1.086 & Pd  & 1.086 \\
Ag  & 1.086 & Cd  & 1.086 \\
In  & 1.034 & Sn  & 1.034 \\
Sb  & 1.034 & Te  & 1.034 \\
I   & 1.034 & Xe  & 1.034 \\
Cs  & 1.294 & Ba  & 1.449 \\
La  & 1.294 & Ce  & 1.319 \\
Pr  & 1.396 & Nd  & 1.449 \\
Pm  & 1.449 & Sm  & 1.449 \\
Eu  & 1.449 & Gd  & 1.449 \\
Tb  & 1.449 & Dy  & 1.449 \\
Ho  & 1.449 & Er  & 1.355 \\
Tm  & 1.449 & Yb  & 1.449 \\
Lu  & 1.449 & Hf  & 1.241 \\
Ta  & 1.294 & W   & 1.189 \\
Re  & 1.189 & Os  & 1.189 \\
Ir  & 1.189 & Pt  & 1.189 \\
Au  & 1.189 & Hg  & 1.189 \\
Tl  & 1.189 & Pb  & 1.189 \\
Bi  & 1.189 &     &       \\
\bottomrule
\end{tabular}
\end{table*}

(II) Exploration: 
The CALYPSO crystal structure prediction method was used to explore crystal configurations under high-pressure conditions.
In iterations 1–102, structures containing three elements with 8–16 atoms were generated; iterations 103–113 expanded this to four elements with 16–32 atoms.
To prevent nonphysical structures with unrealistically short interatomic distances, element-specific safe radii ($r_s$, Table~\ref{tab:safeR})~\cite{mattsson2012lithium} were used. The minimum permissible distance between atoms A and B was defined as
\begin{equation}\label{eq:dmin}
d_{min} =  \frac{r_s(A) + r_s(B)}{1.2}
\end{equation}
In the first three iterations, 100,000 configurations were generated by CALYPSO per cycle. Each configuration was screened using the maximum standard deviation of atomic force predictions (force uncertainty) from the model ensemble; configurations with deviations greater than 1.0 eV/\AA were discarded.
The remaining structures were ranked by force uncertainty, and 5,000 configurations were randomly selected from the top 10\% for subsequent DFT validation.
Iterations 4–37 followed a consistent sampling strategy: CALYPSO generated 100,000 configurations per cycle. These configurations were locally optimized using one model from the ensemble with the L-BFGS algorithm (ASE implementation), allowing up to 20 steps with a force convergence threshold of 20 meV/\AA. Target pressures were randomly sampled from 0–100 GPa. Force uncertainties were evaluated at six key points along each relaxation trajectory (initial, 5th, 10th, 10th-from-last, 5th-from-last, and final configurations). From this pool of key configurations, 20,000 were selected for DFT labeling using the same top-10\%-then-downsampling criterion.
Iterations 38–82 employed the same strategy as iterations 4–37, except the maximum number of optimization steps was increased to 200.
In iterations 83–102, CALYPSO generated 500,000 configurations per cycle. These were optimized using one ensemble model and screened based on enthalpy uncertainty, defined as the standard deviation of the enthalpy predictions from the ensemble. From the top 10\% with the highest enthalpy uncertainty, 20,000 configurations were randomly selected for DFT labeling.
In iterations 103–113, the sampling method was similar to that of iterations 38–82, with the following modifications: the configurations were four-element compounds containing 16–32 atoms, and 5,000 structures were selected from the top 10\% with the highest force uncertainty for DFT labeling.
At the end of iteration 113, the force uncertainty at the 95th percentile (top 5\%) was approximately 0.35 eV/Å.

(III) Labeling: All DFT calculations were performed with the ABACUS code using a plane-wave basis set. The Perdew–Burke–Ernzerhof (PBE) functional~\cite{Perdew1996PBE} within the generalized gradient approximation (GGA) was used to model exchange–correlation effects. A Gaussian smearing width of 0.137 eV was applied. An energy cutoff of 1360 eV and a Monkhorst–Pack k-point grid with a spacing of 0.15 Å\(^{-1}\) were chosen to ensure accurate predictions of energy, force, and virial.
These settings ensure that changes in energy, force, and virial predictions do not exceed 0.001 eV/atom, 0.020 eV/Å, and 0.020 eV/atom, respectively, even when larger energy cutoffs or finer k-point grids are used.
Pseudopotentials were selected from the Optimized Norm-Conserving Vanderbilt (ONCV) set and the PseudoDojo library~\cite{schlipf2015optimization,van2018pseudodojo}.
The complete list of pseudopotentials used is provided in supplementary Table~\ref{tab:pseudopotentials}.
To maintain consistency across the diverse compositional space, we excluded the spin density approximation~\cite{wang1982magnetism}, noncolinear magnetism~\cite{kubler1988density}, and Hubbard U corrections~\cite{kulik2006density}, following the rationale outlined in Refs.~\cite{lopanitsyna2023modeling,mazitov2024surface,mazitov2025petmadlightweightuniversalinteratomic}.
This choice was necessary because applying specialized methods would require element-specific parameterization, introducing inherent inconsistencies in training data for broad transition-metal systems. 
While this limits accuracy for strongly magnetic materials, it ensures a unified treatment of all structures within a single electronic structure framework.

\begin{table*}
\caption{Pseudopotentials used in the DFT-ABACUS calculations.}
\label{tab:pseudopotentials}
\setlength{\tabcolsep}{4pt}
\small
\vspace{-\baselineskip} 
\begin{minipage}[t][][t]{0.48\textwidth}
\vspace{8pt} 
\begin{tabular}{@{}ll@{}}
\toprule
Element & Pseudopotential File \\
\midrule
H & H\_ONCV\_PBE-1.0.upf \\
He & He\_ONCV\_PBE\_FR-1.0.upf \\
Li & Li\_pd\_04\_s-high.UPF \\
Be & Be\_pd\_04\_s-high.UPF \\
B & B.PD03.PBE.UPF \\
C & C.PD04.PBE.UPF \\
N & N\_ONCV\_PBE-1.0.upf \\
O & O.PD04.PBE.UPF \\
F & F.PD03.PBE.UPF \\
Ne & Ne\_ONCV\_PBE-1.0.upf \\
Na & Na-sp.PD04.PBE.UPF \\
Mg & Mg.PD04.PBE.UPF \\
Al & Al.PD04.PBE.UPF \\
Si & Si.PD04.PBE.UPF \\
P & P-sp.PD04.PBE.UPF \\
S & S.PD03.PBE.UPF \\
Cl & Cl.PD04.PBE.UPF \\
Ar & Ar\_ONCV\_PBE-1.0.upf \\
K & K-sp.PD04.PBE.UPF \\
Ca & Ca\_pd\_04\_sp.UPF \\
Sc & Sc-sp.PD04.PBE.UPF \\
Ti & Ti\_pd\_04\_sp.UPF \\
V & V-sp.PD04.PBE.UPF \\
Cr & Cr\_ONCV\_PBE-1.0.upf \\
Mn & Mn-sp.PD04.PBE.UPF \\
Fe & Fe\_ONCV\_PBE-1.2.upf \\
Co & Co\_ONCV\_PBE-1.0.upf \\
Ni & Ni-sp.PD04.PBE.UPF \\
Cu & Cu\_sg15\_1.2\_.UPF \\
Zn & Zn\_pd\_04\_.UPF \\
Ga & Ga\_pd\_04\_d.UPF \\
Ge & Ge\_pd\_04\_d.UPF \\
As & As-d.PD04.PBE.UPF \\
Se & Se.upf \\
Br & Br.PD03.PBE.UPF \\
Kr & Kr\_ONCV\_PBE-1.2.upf \\
Rb & Rb-sp.PD04.PBE.UPF \\
Sr & Sr-sp.PD04.PBE.UPF \\
Y & Y-sp.PD04.PBE.UPF \\
Zr & Zr-sp.PD04.PBE.UPF \\
Nb & Nb-sp.PD04.PBE.UPF \\
Mo & Mo-sp.PD04.PBE.UPF \\
Tc & Tc\_ONCV\_PBE-1.0.upf \\
Ru & Ru-sp.PD04.PBE.UPF \\
Rh & Rh-sp.PD04.PBE.UPF \\
Pd & Pd-sp.PD04.PBE.UPF \\
Ag & Ag\_ONCV\_PBE\_FR-1.0.upf \\
Cd & Cd\_ONCV\_PBE-1.2.upf \\
In & In\_pd\_04\_d.UPF \\
Sn & Sn-d.PD04.PBE.UPF \\
Sb & Sb.PD03.PBE.UPF \\
Te & Te.PD04.PBE.UPF \\
\bottomrule
\end{tabular}
\end{minipage}
\hfill
\begin{minipage}[t][][t]{0.48\textwidth}
\vspace{8pt} 
\begin{tabular}{@{}ll@{}}
\toprule
Element & Pseudopotential \\
\midrule
I & I.upf \\
Xe & Xe.upf \\
Cs & Cs.upf \\
Ba & Ba.upf \\
La & La\_pd\_04\_sp.UPF \\
Ce & Ce\_pd\_04\_3+\_f--core.UPF \\
Pr & Pr3+\_f--core.PD04.PBE.UPF \\
Nd & Nd3+\_f--core.PD04.PBE.UPF \\
Pm & Pm3+\_f--core-icmod1.PD04.PBE.UPF \\
Sm & Sm3+\_f--core.PD04.PBE.UPF \\
Eu & Eu3+\_f--core-icmod1.PD04.PBE.UPF \\
Gd & Gd\_pd\_04\_3+\_f--core.UPF \\
Tb & Tb\_pd\_04\_3+\_f--core.UPF \\
Dy & Dy\_pd\_04\_3+\_f--core-icmod1.UPF \\
Ho & Ho3+\_f--core.PD04.PBE.UPF \\
Er & Er3+\_f--core.PD04.PBE.UPF \\
Tm & Tm\_pd\_04\_3+\_f--core-icmod1.UPF \\
Yb & Yb3+\_f--core-icmod1.PD04.PBE.UPF \\
Lu & Lu3+\_f--core.PD04.PBE.UPF \\
Hf & Hf\_pd\_04\_sp.UPF \\
Ta & Ta\_pd\_04\_sp.UPF \\
W & W\_ONCV\_PBE-1.1.upf \\
Re & Re-sp.PD04.PBE.UPF \\
Os & Os-sp.PD04.PBE.UPF \\
Ir & Ir-sp.PD04.PBE.UPF \\
Pt & Pt\_pd\_04\_sp.UPF \\
Au & Au\_ONCV\_PBE-1.0.upf \\
Hg & Hg\_ONCV\_PBE-1.2.upf \\
Tl & Tl\_ONCV\_PBE-1.0.upf \\
Pb & Pb\_pd\_03\_.UPF \\
Bi & Bi-spd-high.PD04.PBE.UPF \\
\bottomrule
\end{tabular}
\end{minipage}
\end{table*}

\subsection{OpenCSP model training schemes}\label{sec:training}

The OpenCSP models, OpenCSP-L6, OpenCSP-L12, and OpenCSP-L24, are based on the DPA3 architecture, which employs graph neural networks constructed on an atom-bond graph (\(G^{(1)}\)) and a bond-angle graph (\(G^{(2)}\)). 
The bond-angle graph is derived from the atom-bond graph through a line graph transformation.
The vertex and edge feature dimensions of \(G^{(1)}\) are 128 and 64, respectively, while the edge feature dimension of \(G^{(2)}\) is 32.
The depths of OpenCSP-L6, OpenCSP-L12, and OpenCSP-L24 are 6, 12, and 24 layers, respectively.
The cutoff radius for \(G^{(1)}\) is set to 6.0 Å, with a smooth switching function applied between 3 and 6 Å. Similarly, the cutoff radius for \(G^{(2)}\) is set to 4.0 Å, with a smooth switching function applied between 2 and 4 Å.
The learning rate was adjusted using an exponential decay strategy throughout the training process.
All three models underwent training in two stages, with each stage consisting of 500,000 steps.
In the first stage, the learning rate decayed exponentially from \(1\times10^{-3}\) to \(1\times10^{-5}\). 
Concurrently, the prefactors of the mean square loss terms for energy, force, and virial evolved linearly from 0.2, 100, and 0.02 to 20, 20, and 1.0, respectively, as the learning rate decreased. 
In the second stage, the learning rate decayed exponentially from \(1\times10^{-4}\) to \(1\times10^{-5}\).
During this stage, the prefactors for the energy, force, and virial loss terms remained constant at 15, 1.0, and 2.5, respectively.
The batch sizes, defined as the number of configurations in a minibatch, were determined such that the product of the number of atoms per configuration and the number of configurations in the batch was approximately 4096 for the OpenCSP-L6 and OpenCSP-L12 models. 
For the OpenCSP-L24 model, this product was adjusted to 1536 to accommodate its extensive memory requirements.

\section{Author Contributions}
H.W. and J.L. conceived and designed the research. 
YN.W., XY. W., ZY. W. and J. W. performed the calculations and data analysis. 
YN.W., XY. W., and ZY. W. wrote the original draft of the manuscript. 
All authors discussed the results, reviewed, edited the manuscript, and approved the final manuscript.
\section{Acknowledgements}
The work of Han Wang is supported by the National Key R\&D Program of China (Grant No.~2022YFA1004300).
J.L. wants to thank the National Natural Science Foundation of China (Grants Nos. 12034009, 12374005). 
XY.W wants to thank the National Natural Science Foundation of China (Grants Nos. 12404004).
The work of ZY. W. is supported by the Postdoctoral Fellowship Program of CPSF under Grant Number GZC20252243.
\section{Competing interests}
All authors declare no financial or non-financial competing interests. 
\section{Computer code}
CALYPSO code is free for academic use, by registering at http://www.calypso.cn. 
ASE, DP-GEN, and DeepMD-kit are free and open source codes available at https://ase-lib.org/ and https://deepmodeling.com, respectively. 
The other scripts are available from the authors upon request.
\section{Data availability}
The OpenCSP datasets and models are available in the AIS Square repository, https://www.aissquare.com/models?page=1\&type=models, https://www.aissquare.com/datasets?page=1\&type=datasets.


\providecommand{\noopsort}[1]{}\providecommand{\singleletter}[1]{#1}%
\bibliographystyle{apsrev4-2}

\end{document}